\documentclass[10pt,a4paper]{article}

\usepackage{arxiv}
\usepackage[utf8]{inputenc}
\usepackage[T1]{fontenc}
\usepackage{amsmath,amssymb,amsfonts,latexsym}
\usepackage{mathtools}
\usepackage{graphicx}
\usepackage{booktabs}
\usepackage{xcolor}
\usepackage{url}
\usepackage{float}
\usepackage{placeins}
\usepackage[numbers,sort&compress]{natbib}
\usepackage{hyperref}
\usepackage[nameinlink,capitalize]{cleveref}

\allowdisplaybreaks

\graphicspath{{figures/}}
\newcommand{\dd}{\,\mathrm{d}}

\DeclareMathOperator{\Dc}{{}^{C}\!D}
\newcommand{\GD}{\mathbb{D}}
\newcommand{\GI}{\mathbb{I}}
\newcommand{\Ir}{I}

\title{How much data does a memory kernel need? Measurement requirements for fractional models of battery and corrosion degradation}

\author{%
  Conrard Giresse Tetsassi Feugmo\textsuperscript{1,2,*}\\[4pt]
  \small\textsuperscript{1}Department of Chemistry, University of Waterloo,
  Waterloo, ON N2L 3G1, Canada\\
  \small\textsuperscript{2}Department of Physics and Astronomy, University of Waterloo,
  Waterloo, ON N2L 3G1, Canada\\[2pt]
  \small\textsuperscript{*}Author to whom any correspondence should be addressed.
  \href{mailto:cgtetsas@uwaterloo.ca}{cgtetsas@uwaterloo.ca}
}

\date{}

\begin{document}
\maketitle
\begin{abstract}
How many decades of observation does a memory kernel need before a fit tells us something? For a distributed order, the answer is $3.1$ decades, and no standard degradation protocol reaches it. Battery capacity fade and corrosion film growth depend on their past, so they are often described with general fractional calculus (GFC). In GFC the memory is an arbitrary Sonine kernel rather than a single power law. Fitting such a kernel to a measured aging curve is appealing, but it is unclear what the fit returns. We test this on ground truth from two mechanistic models that contain no fractional operator, and on de Levie's closed-form porous electrode, whose high-frequency order is exactly $\tfrac12$. The estimator returns $0.500$ for that electrode. What one curve can recover depends on the question. A single aging curve gives an effective order $\alpha$, which holds to $5\%$ noise when a weak formulation is used, because derivative matching breaks down under noise. The tempered pair $(\alpha,\lambda)$ can be separated only with several Arrhenius conditions, and the joint error drops from $(|\Delta\alpha|,|\Delta\lambda|)=(0.25,0.43)$ at one condition to $(0.04,0.07)$ at four. A distributed order needs $3.1$ decades of range to be resolved to $90\%$. Below that, a two-order memory collapses onto one intermediate order that fits the data better than the true spectrum. Saturating passivation is the exception: a general kernel raises $R^2$ from $0.862$ to $0.998$. Across twenty cells from two public datasets, observation windows span $1.23$ to $2.24$ decades. The $3.1$-decade figure therefore works as a design rule. It says what to measure before fitting, not only what cannot be fitted.
\end{abstract}

\section{Introduction}\label{sec:1}
\setcounter{equation}{0}

Capacity fade in lithium-ion cells and passive-film growth in corrosion follow non-integer power laws and change regime over life. Such behaviour is a signature of memory and anomalous transport~\cite{MetzlerKlafter2000}, and fractional-order models have been fitted to lithium-ion battery impedance and capacity data~\cite{WangChenLiao2019,TianXiong2020,WangLiChen2020}. Fractional calculus defines derivatives of non-integer order $\alpha$ through history-weighted convolutions~\cite{Podlubny1999,Kilbas2006}. Its general form, general fractional calculus (GFC), replaces the power-law kernel by an arbitrary Sonine kernel while a fundamental theorem of calculus still holds~\cite{Sonine1884,Kochubei2011,Luchko2021,Tarasov2024}.

Inverting a measured aging curve for such a kernel is attractive, because an arbitrary kernel can encode a distribution of underlying processes. The same freedom causes a difficulty. A model with more freedom can fit the data better than the process that generated them. Fit a two-order memory, orders $0.3$ and $0.8$, to a simulated aging curve spanning two decades of time, and the fit returns a single intermediate order near $0.42$. That order reconstructs the curve at $R^{2}=0.9989$, better than the two-order spectrum that produced it, and the residuals do not reveal the failure.

Existing analytic results establish when a recovery is unique and how its error behaves in the limit (Section~\ref{sec:rw}). They do not state how many decades of observation a protocol must span before the recovery carries information, which is the quantity an experimentalist sets. This paper supplies that quantity.

This paper makes three points. The first concerns how much data is needed. Resolving a distributed order to $90\%$ takes $3.1$ decades of observation. The observation windows of twenty cells from two public datasets span $1.23$ to $2.24$ decades, so standard degradation protocols fall short of the requirement by $0.9$ to $1.9$ decades. The second concerns what a single curve can recover. An effective order $\alpha$ is fixed by one aging curve, and it holds to $5\%$ noise under a weak formulation where derivative matching collapses. The tempered pair $(\alpha,\lambda)$ is identifiable only across several Arrhenius conditions, with the joint error falling from $(0.25,0.43)$ at one condition to $(0.04,0.07)$ at four. Saturating passivation is the exception, where a general kernel raises the reconstruction $R^{2}$ from $0.862$ to $0.998$. The third point concerns the estimator itself. It is checked against ground truth that contains no fractional operator: two nonlinear moving-boundary models, and de Levie's porous electrode, whose high-frequency order is exactly $\tfrac12$. The estimator returns $0.500$ for that electrode.

The rest of this paper is organised as follows. Section~\ref{sec:rw} reviews the inverse-problem literature. Section~\ref{sec:2} sets out the operators, the kernel families, the mechanistic ground truth and the inverse problem. Section~\ref{sec:3} reports what is recoverable at each level and measures the dynamic-range boundary against published cells. Section~\ref{sec:4} assembles the identifiability map, and Section~\ref{sec:5} concludes. The method recovers a reduced, effective equation of motion. It does not recover the governing partial differential equation, a spatial field or the microscopic rate constants. Nothing here argues that general fractional models are worse than power laws; on a single curve they are usually indistinguishable from them.

\section{Related work}\label{sec:rw}

Recovering coefficients, orders or kernels from data has a large analytic literature. The present paper asks a different question, so the two are complementary. Recent work in this journal studies time-fractional inverse problems. It covers source identification~\cite{Zeng2026}, recovery of a coefficient that depends on time~\cite{BayrakDemir2024}, a driving term of fractional Brownian type~\cite{LeiFanLi2024}, and a multiplicative noise term~\cite{YangMaWo2026}. Cheng et~al.~\cite{Cheng2009} showed that the order of a time-fractional diffusion equation can be determined uniquely. Jin and Rundell~\cite{JinRundell2015} reviewed this result and the wider anomalous-diffusion inverse problem. Liu and Yamamoto~\cite{LiuYamamoto2023} extended it to several orders at once, and showed that short-time behaviour separates the orders of a multi-term memory. Jin and Kian showed recovery in an unknown medium for a single order~\cite{JinKian2022}, and, most relevant here, for a \emph{distributed} order~\cite{JinKian2023}. Cen et~al.~\cite{CenJinLiuZhou2023} asked how much a limited measurement can determine, and recovered several parameters of a subdiffusion problem from one lateral boundary measurement. Zhang and Zhou~\cite{ZhangZhou2023} studied the stability of this class of ill-posed problems and its numerical treatment. Kaltenbacher and Rundell treated attenuation and damping terms~\cite{KaltenbacherRundell2021,KaltenbacherRundell2022}, and competing coefficients in a nonlinear setting~\cite{KaltenbacherRundell2026}. Data-driven equation discovery has moved from sparse regression~\cite{Brunton2016} to fractional physics-informed networks~\cite{Pang2019,Gulian2019}, and it has since been extended to non-integer operators, including distributed-order equations~\cite{Momeni2025}.

This work shows when a recovery is unique and how its error behaves in the limit. It does not say how many decades a protocol must span before the recovery is meaningful. An experimenter needs that number, and supplying it is what this paper does.

\section{General fractional operators, kernel families, and the inverse problem}\label{sec:2}
\setcounter{equation}{0}

This section can be read on its own. It defines the operators and the memory kernels, then the mechanistic models used as ground truth, then the numerical method, and last the inverse problem: what we try to recover and how. Readers who only want the results can skip to Section~\ref{sec:3}, but the claims there depend on the definitions given here.

\subsection{From integer to fractional derivatives}\label{sec:2.1}

We start with the Riemann--Liouville fractional integral of order $\alpha>0$. It averages the past of $u$ with a power-law weight,
\begin{equation}
\big(\Ir^{\alpha}u\big)(t)=\frac{1}{\Gamma(\alpha)}\int_0^{t}(t-\tau)^{\alpha-1}\,u(\tau)\dd\tau
=\big(\kappa_\alpha * u\big)(t),
\qquad
\kappa_\alpha(t)=\frac{t^{\alpha-1}}{\Gamma(\alpha)},
\label{eq:RLint}
\end{equation}
Here $\Gamma$ is the gamma function. The symbol $*$ is the convolution, a weighted sum over the past, and $\kappa_\alpha$ is the power-law weight. When $\alpha=1$ this is ordinary integration. For other values of $\alpha$, every past value counts, although older values count for less. This is what makes the operator \emph{nonlocal} and gives it ``memory.'' Differentiating once, and using the weight of order $1-\alpha$, gives the Caputo derivative for $\alpha$ between 0 and 1,
\begin{equation}
\big(\Dc^{\alpha}u\big)(t)
=\frac{1}{\Gamma(1-\alpha)}\int_0^{t}(t-\tau)^{-\alpha}\,u'(\tau)\dd\tau
=\big(\kappa_{1-\alpha}*u'\big)(t).
\label{eq:caputo}
\end{equation}
The Caputo form differentiates first and then averages. A constant has zero derivative under this form, and ordinary starting values can be used~\cite{Podlubny1999,Diethelm2010}. That makes it a natural choice for degradation problems that start from a known state. Other definitions exist, and they are not equivalent. The Riemann--Liouville form averages first and differentiates second. The Grünwald--Letnikov form is a limit of fractional differences. The Atangana--Baleanu form uses a kernel with no singularity~\cite{AtanganaBaleanu2016}, and some researchers dispute whether that kernel is well behaved~\cite{Hanyga2020}. We use the Caputo form throughout~\cite{Caputo1967}. A singular-kernel operator and its numerical treatment are studied in~\cite{Odibat2024}. The standard
references for this material are the monographs of Oldham and
Spanier~\cite{OldhamSpanier1974}, Miller and Ross~\cite{MillerRoss1993}, Podlubny~\cite{Podlubny1999}, Kilbas, Srivastava and
Trujillo~\cite{Kilbas2006} and Diethelm~\cite{Diethelm2010}, with the historical account
of Ross~\cite{Ross1977} and the geometric reading of Tarasov~\cite{Tarasov2016geo};
applications across physics, viscoelasticity, bioengineering and dynamics are surveyed
in~\cite{Hilfer2000,Mainardi2010,Magin2006,Herrmann2011,Uchaikin2013,Tarasov2010,%
BagleyTorvik1983,Sun2018,Kiryakova1994,TarasovTarasova2021}, and the field's open
problems are collected in~\cite{DiethelmTrends2022}. A power function is the standard test case,
\begin{equation}
\Dc^{\alpha}\big[t^{\beta}\big]=\frac{\Gamma(\beta+1)}{\Gamma(\beta+1-\alpha)}\,t^{\beta-\alpha},
\qquad\text{in particular}\qquad
\Dc^{\alpha}\big[t^{\alpha}\big]=\Gamma(\alpha+1),
\label{eq:monomial}
\end{equation}
We use it to check the numerical method below, where the relative error is about $\sim\!10^{-6}$.

\subsection{The general (Sonine-kernel) fractional calculus}\label{sec:2.2}

The power law in ~\eqref{eq:RLint}--\eqref{eq:caputo} is only one choice of memory. General fractional calculus allows any pair of kernels instead. Let $K$ and $M$ be two kernels that form a \emph{Sonine pair}. This means they satisfy the Sonine condition~\cite{Sonine1884,Luchko2021}
\begin{equation}
(K*M)(t)=\int_0^{t}K(t-\tau)M(\tau)\dd\tau = 1 \quad (t>0),
\qquad\Longleftrightarrow\qquad
\widehat K(s)\,\widehat M(s)=\frac1s,
\label{eq:sonin}
\end{equation}
The right-hand form is the same condition after a Laplace transform, which we mark with a hat ($\widehat{\phantom{K}}$). For such a pair, the general fractional integral and derivative are the convolutions
\begin{equation}
\big(\GI^{(K)}u\big)(t)=(K*u)(t),
\qquad
\big(\GD^{(M)}u\big)(t)=(M*u')(t).
\label{eq:GFC}
\end{equation}
The Sonine condition is what makes the two operators inverses of each other. The \emph{fundamental theorem of general fractional calculus}~\cite{Kochubei2011,Luchko2021op} says that
\begin{equation}
\big(\GI^{(K)}\GD^{(M)}u\big)(t)=u(t)-u(0),
\label{eq:ftc}
\end{equation}
This extends the ordinary result $\int_0^t u'=u(t)-u(0)$. The Caputo operator is the special case $M=\kappa_{1-\alpha}$, $K=\kappa_{\alpha}$. Then $\widehat K\widehat M=s^{-\alpha}s^{\alpha-1}=1/s$. To keep the memory physical, we require $M$ to be completely monotone. Its influence then shrinks as the past gets older. By Bernstein's theorem, such kernels are exactly the non-negative mixtures of exponentials, $M(t)=\int_0^\infty e^{-rt}\dd\nu(r)$. Our identification method uses this property. Complete monotonicity is not enough by itself. A kernel can be completely monotone without a singularity at the origin. A Sonine pair, however, must have at least one kernel that is singular at the origin, as $\kappa_{1-\alpha}$ is. We use only kernels that meet this stronger condition.

This calculus grew from Sonine's convolution condition~\cite{Sonine1884} and was placed on
a modern footing by Kochubei~\cite{Kochubei2011,KochubeiKondratiev2019} and by a series of works of Luchko~\cite{Luchko2021,Luchko2021op} and Luchko and Yamamoto~\cite{LuchkoYamamoto2016}, and a survey in Kochubei and
Luchko~\cite{KochubeiLuchko2019}. Its physical motivation is anomalous transport: the
continuous-time random walk of Scher and Montroll~\cite{ScherMontroll1975} and the
fractional-kinetics continuum built on
it~\cite{MetzlerKlafter2000,MetzlerKlafter2004,SokolovKlafter2002,Zaslavsky2002}, whose
fundamental solutions~\cite{MainardiLuchkoPagnini2001,GorenfloMainardi2003} and
Mittag-Leffler relaxation
functions~\cite{GorenfloMainardiMLF2014,MainardiGorenflo2000} recur throughout what
follows.

\subsection{Kernel families}\label{sec:2.3}

Within ~\eqref{eq:GFC} we use four kernels, from the simplest to kernels that are not power laws.
\emph{(i) Single order (power law):} $M=\kappa_{1-\alpha}$, one parameter $\alpha$.
\emph{(ii) Tempered power law:} an exponential cutoff at rate $\lambda\ge0$,
\begin{equation}
M_{\alpha,\lambda}(t)=\frac{t^{-\alpha}e^{-\lambda t}}{\Gamma(1-\alpha)},
\label{eq:tempered}
\end{equation}
It acts like a fractional derivative at short times and fades exponentially at long times. This is the signature of a memory that saturates. \emph{(iii) Distributed order:} several orders act at the same time. They are weighted by a spectrum $w(\alpha)\ge0$, $\int_0^1 w=1$,
\begin{equation}
\big(\GD^{(w)}u\big)(t)=\int_0^1 w(\alpha)\,\big(\Dc^{\alpha}u\big)(t)\dd\alpha,
\label{eq:distorder}
\end{equation}
whose kernel is the corresponding mixture of power laws. \emph{(iv) Time-varying order:} the order itself changes with time, $\alpha=\alpha(t)$. This gives a change of regime over the life of the cell, for example a fast early phase that bends into a slower one.

The tempered kernel forms a valid Sonine pair, but its partner is not the tempered power law itself. Since $\widehat M_{\alpha,\lambda}(s)=(s+\lambda)^{\alpha-1}$, the Sonine condition ~\eqref{eq:sonin} then forces
\begin{equation}
\begin{aligned}
\widehat K_{\alpha,\lambda}(s)&=\frac{1}{s\,\widehat M_{\alpha,\lambda}(s)}
=\frac{(s+\lambda)^{1-\alpha}}{s},
\qquad\text{that is}\\[2pt]
K_{\alpha,\lambda}(t)&=\frac{t^{\alpha-1}e^{-\lambda t}}{\Gamma(\alpha)}
+\frac{\lambda}{\Gamma(\alpha)}\int_0^{t}\tau^{\alpha-1}e^{-\lambda\tau}\dd\tau .
\end{aligned}
\label{eq:temperedpair}
\end{equation}
The tempered power law $\kappa_{\alpha}^{\lambda}(t)=t^{\alpha-1}e^{-\lambda t}/\Gamma(\alpha)$ alone is \emph{not} the partner. Its transform is $(s+\lambda)^{-\alpha}$, so
$\widehat\kappa_{\alpha}^{\lambda}\widehat M_{\alpha,\lambda}=(s+\lambda)^{-1}\neq 1/s$
for $\lambda>0$. The second term in ~\eqref{eq:temperedpair} restores the Sonine condition. Because $(s+\lambda)^{1-\alpha}/s$ is a Stieltjes function, $K_{\alpha,\lambda}$ is completely monotone. The pair therefore belongs to the Kochubei class. This is the worked example in~\cite{Luchko2021}. We treat families (iii) and (iv) as ways to generate memory: non-negative mixtures and slow changes of admissible power-law kernels. They are not single closed-form Sonine operators. They are the non-power-law cases we test for identifiability.

\subsection{Where the order and the tempering come from}\label{sec:2.3b}

Families (i) and (ii) are not arbitrary. The order $\alpha$ and the tempering rate $\lambda$ both come from the physics of trap-limited transport. Knowing their origin turns two fitting constants into measurable material properties. Consider an ion hopping on a
disordered lattice. An ion escapes from a site of depth $E$ by thermal activation,
$\tau_w=\nu_0^{-1}e^{E/k_BT}$ with attempt frequency $\nu_0$. Assume the trap depths follow a standard exponential distribution, $p(E)=E_0^{-1}e^{-E/E_0}$ for $E\ge0$, where $E_0$ measures the disorder. Changing variables from $E$ to the waiting time $\tau_w$, with $\mathrm{d}E/\mathrm{d}\tau_w=k_BT/\tau_w$, gives the waiting-time density
\begin{equation}
\psi(\tau_w)=p\big(E(\tau_w)\big)\,\frac{\mathrm{d}E}{\mathrm{d}\tau_w}
=\alpha\,\nu_0^{-\alpha}\,\tau_w^{-1-\alpha},
\qquad
\boxed{\;\alpha=\frac{k_BT}{E_0}\;}
\label{eq:alphaphys}
\end{equation}
An exponential spread of \emph{energies} becomes a power-law spread of \emph{times}. For
$\alpha<1$ the mean waiting time $\int\tau_w\psi(\tau_w)\dd\tau_w$ diverges, so no finite
diffusion coefficient exists; the continuous-time random walk then has the time-fractional
equation ~\eqref{eq:caputo} as its continuum limit
\cite{ScherMontroll1975,MetzlerKlafter2000}. The order is therefore the ratio of thermal
energy to disorder width rather than a free exponent, and ~\eqref{eq:alphaphys} predicts that
$\alpha$ grows \emph{linearly} with temperature.

Equation~\eqref{eq:alphaphys} assumes traps of unbounded depth. A real material has a
deepest site $E_{\max}$, hence a longest residence time
$\tau_{w,\max}=\nu_0^{-1}e^{E_{\max}/k_BT}$. Beyond $\tau_{w,\max}$ every ion has escaped even
the worst trap and transport reverts to ordinary diffusion, so the power-law tail of
$\psi$ is cut off. That cutoff is precisely the exponential factor of the tempered kernel
\eqref{eq:tempered}, with
\begin{equation}
\lambda\simeq\tau_{w,\max}^{-1}=\nu_0\,e^{-E_{\max}/k_BT}.
\label{eq:lambdaphys}
\end{equation}
Tempering is thus the mathematical image of bounded disorder, and a passivating or
saturating curve is the signature of a material operating past its crossover time.

\paragraph{A worked example.} The two parameters behave differently enough that it helps
to put numbers to them. Take a disordered ceramic electrolyte with disorder
width $E_0=0.05$\,eV, deepest trap $E_{\max}=0.5$\,eV, and attempt frequency
$\nu_0=10^{13}\,$s$^{-1}$. Using $k_B=8.617\times10^{-5}$\,eV\,K$^{-1}$,
\eqref{eq:alphaphys} and ~\eqref{eq:lambdaphys} give Table~\ref{tab:worked}.

\begin{table}[!ht]
\begin{center}
\caption{The order and the tempering rate of a disordered ceramic electrolyte with
disorder width $E_0=0.05$\,eV, deepest trap $E_{\max}=0.5$\,eV and attempt frequency
$\nu_0=10^{13}\,$s$^{-1}$, from ~\eqref{eq:alphaphys} and ~\eqref{eq:lambdaphys}.}
\label{tab:worked}
\begin{tabular}{lccc}
\toprule
$T$ (K) & $k_BT$ (eV) & $\alpha=k_BT/E_0$ & $\lambda=\nu_0e^{-E_{\max}/k_BT}$ (s$^{-1}$)\\
\midrule
$250$ & $0.0215$ & $0.431$ & $8.3\times10^{2}$ \\
$300$ & $0.0259$ & $0.517$ & $4.0\times10^{4}$ \\
$350$ & $0.0302$ & $0.603$ & $6.3\times10^{5}$ \\
\bottomrule
\end{tabular}
\end{center}
\end{table}

Over the same $100$\,K the order $\alpha$ changes by $40\%$, strictly linearly, while the
tempering rate $\lambda$ changes by a factor of about $760$, exponentially. At $300$\,K the
crossover time is $\tau_{w,\max}=1/\lambda\approx25\,\mu$s: below it transport is anomalous
with $\alpha\approx0.5$, above it ordinary. This is the concrete content of
``$\alpha\approx\tfrac12$ looks like Fickian diffusion'': here it is the statement
$k_BT\approx E_0/2$, not an empirical coincidence.

This example also explains how the identification problem of Section~\ref{sec:2.7} behaves. Because $\alpha$ is linear in $T$, and $\lambda$ follows an Arrhenius law in $1/T$, a change in temperature moves the two parameters in different directions. They change by $40\%$ and by nearly three orders of magnitude. Measuring at several temperatures therefore separates them. More data at one temperature would not. This is why the joint fit in Section~\ref{sec:3} works where a single curve fails.

\subsection{Operators with respect to another function}\label{sec:2.4}

Aging follows an internal clock, not laboratory time. Temperature and load speed it up or slow it down. Tarasov's parametric calculus~\cite{Tarasov2024} handles this with an internal clock $g(t)$. The clock increases strictly and starts at $g(0)=0$. Time is changed with the substitution operator $\big(Q_g u\big)(t)=u\!\big(g(t)\big)$. The operator ``with respect to $g$'' is the ordinary operator, wrapped by $Q_g$:
\begin{equation}
\GD^{(M)}_{g}=Q_g\,\GD^{(M)}\,Q_g^{-1},
\label{eq:Qg}
\end{equation}
In words: move the data to the internal clock (using $g^{-1}$), apply the ordinary operator, then move back to laboratory time (using $g$). An Arrhenius clock, $g(t)=k(T)\,t$ with $k(T)=A\,e^{-E_a/(R_g T)}$ (where $A$ is the pre-exponential factor, $E_a$ the activation energy, $R_g$ the gas constant and $T$ the absolute temperature), makes different temperatures the \emph{same} process run at different speeds, with the \emph{same} kernel. The multi-condition method relies on this. Section~\ref{sec:2.7} asks over what range this assumption is a good approximation.

\subsection{Independent mechanistic ground truth}\label{sec:2.5}

To avoid circular reasoning, we also generate data from physics that has no fractional operator. We then ask whether the memory description can recover it. We solve two moving-boundary models implicitly, using the Jacobian-free Newton--Krylov method \cite{KnollKeyes2004}. The method uses GMRES with automatic differentiation, so the Jacobian is never formed. The first is a reaction-diffusion model of growth in the solid electrolyte interphase, a thin film on a battery electrode. A species diffuses through the film, and the film thickness $L(t)$ grows as that species is used up. If diffusion limits the growth, $L\!\sim\!\sqrt{t}$, and the capacity loss has effective order $\alpha\!\approx\!1/2$,
a signature usually associated with passivation~\cite{PinsonBazant2013}. The second is the Point Defect Model (PDM) of a passive oxide film~\cite{Macdonald1992}. Oxide grows under a strong electric field and also dissolves chemically. The two balance at a steady thickness, which gives a memory that saturates. Here ``bounded'' describes the physics. When we fit a general kernel to the saturating curve, the best description is one sharp intermediate order. It is not a mix of two orders, nor a simple tempered pair, as Section~\ref{sec:3} reports. The two mechanistic models are the reference processes; the fractional description is what we fit to their output.

In other tests, we run the forward model with a known kernel and add realistic measurement effects. These include check-ups at irregular times, noise, slow sensor drift, occasional outliers, rounding, and differences from cell to cell. The known answer lets us score how well recovery works.

\subsection{Numerical discretisation}\label{sec:2.6}

We compute the Caputo derivative ~\eqref{eq:caputo} on a time grid $t_n=n h$ with the L1 scheme~\cite{LinXu2007,Diethelm2010}. This is one of several methods for fractional operators~\cite{Lubich1986,YusteAcedo2005,Garrappa2018,LiZeng2015,DiethelmFordFreed2002,DiethelmFord2002,Baleanu2012,GorenfloMainardi1997}. It replaces $u'$ with straight-line differences, and it integrates the power-law weight exactly on each step,
\begin{equation}
\big(\Dc^{\alpha}u\big)(t_n)\approx
\frac{h^{-\alpha}}{\Gamma(2-\alpha)}\sum_{j=0}^{n-1} b_j\,\big(u_{n-j}-u_{n-j-1}\big),
\qquad
b_j=(j+1)^{1-\alpha}-j^{1-\alpha}.
\label{eq:L1}
\end{equation}
The weights $b_j$ decay slowly. This is the discrete form of memory: every past step still counts. General kernels use the same method, with $M$ in place of the power law. Responses over many decades of time are computed in the Laplace domain and then inverted with the fixed-Talbot method~\cite{AbateValko2004,AbateWhitt2006}. That method covers a range of time scales that a single grid cannot. All operators are written in pure JAX and released as the \texttt{gfckit} package.

\subsection{The inverse problem: what we recover, and how}\label{sec:2.7}

\begin{figure}[t]
\begin{center}
\includegraphics[width=\linewidth]{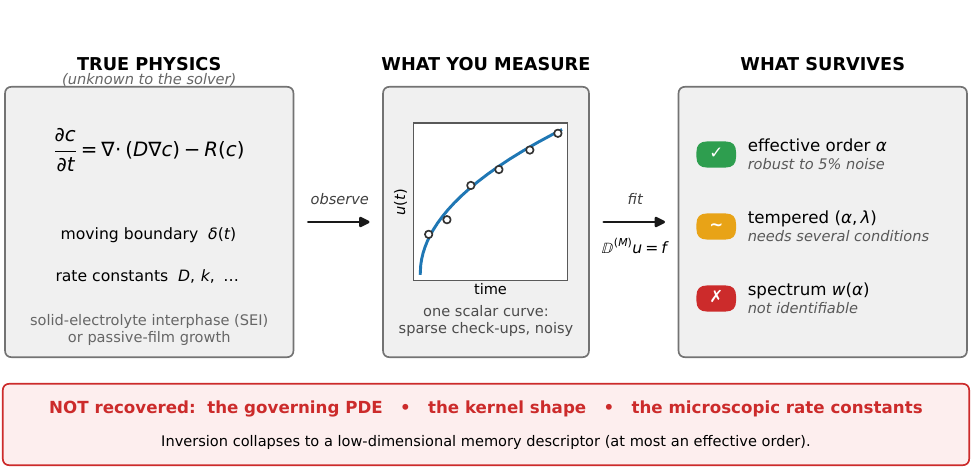}
\caption{What the inverse problem recovers. The true physics (a moving-boundary
reaction-diffusion PDE for solid-electrolyte-interphase (SEI) or passive-film growth,
with its rate constants, left) is never inverted; from a single aging curve, sampled at
sparse check-ups and carrying measurement scatter (center; the curve is the analytic
$\sqrt{t}$ uptake law and the open markers are the check-ups), we fit a fixed-form
fractional evolution equation $\GD^{(M)}u=f$ and recover only a low-dimensional memory
descriptor (right): the effective order $\alpha$ robustly, the tempered pair
$(\alpha,\lambda)$ only across several Arrhenius conditions, and the full order spectrum
$w(\alpha)$ not at all. The governing PDE, the kernel shape, and the microscopic rate
constants are not recovered. This figure is a schematic, not a fitted result.}
\label{fig:recover}
\end{center}
\end{figure}

We first say exactly what the inverse problem is, since it is easy to overstate. Here it recovers only the memory kernel of a fixed-form fractional equation. It does not recover the governing PDE, the microscopic rate constants of the chemistry, or any spatial field. Concretely, we \emph{fix the form} of the model. It is a general fractional evolution equation in time, and its only unknown is the memory kernel $M$,
\begin{equation}
\big(\GD^{(M)}u\big)(t)=f(t),
\label{eq:inverse}
\end{equation}
Here $f$ is a simple, known forcing, such as a constant or slow rate. We ask which features of $M$ a single curve $u(t)$ determines. In other words, we recover a \emph{reduced, effective} equation of motion. It is not the true governing equation, and it is not the full kernel shape. The unknowns, in order of increasing difficulty, are: (a) a single effective order $\alpha$ (the kernel
is the pure power law $\kappa_{1-\alpha}$); (b) the two tempered parameters
$(\alpha,\lambda)$ of ~\eqref{eq:tempered}; and (c) the distributed-order spectrum
$w(\alpha)$ of ~\eqref{eq:distorder}. From a sampled, noisy curve $u$, we map how well each of the three can be recovered: (a) well, (b) only with several conditions, and (c) hardly at all (Fig.~\ref{fig:recover}).

\emph{(a) The effective order $\alpha$.} We use two estimators. The strong form picks the order that makes the derivative as constant as possible,
$\hat\alpha=\arg\min_\alpha \operatorname{Var}_t\big[\Dc^{\alpha}u\big]$, where $\operatorname{Var}_t$ is the sample variance over the time grid. The method is exact, but it is sensitive to noise, because ~\eqref{eq:L1} differentiates the data. The weak (integral) form fits the integrated curve, $u(t)\approx c_1\,t^{\alpha}+c_0$, by nonlinear least squares. Integration smooths noise, so this is the estimator we trust.

\emph{(b) The tempered pair $(\alpha,\lambda)$.} On a single curve they are correlated, because a steeper power law with an earlier cutoff can look like a shallower one. So we fit them \emph{jointly} across Arrhenius conditions. The substitution operator ~\eqref{eq:Qg} turns a change in temperature into a simple rescaling of time. So we use one shared short-time exponent $\alpha$, while the cutoff $\lambda$ scales with the clock $k(T)$. This shared structure removes the degeneracy. This is an approximation, and the same physics sets its limits. If $\alpha=k_BT/E_0$ as in \eqref{eq:alphaphys}, then $\alpha$ itself drifts with temperature, by roughly $40\%$
over the $100$~K of the worked example in Section~\ref{sec:2.3b}. So sharing $\alpha$ is justified only over a narrow temperature range, where the drift is small compared with the fitted uncertainty. The recovery reported below should be read as an average exponent over that range, not as a constant that holds at every temperature. A natural next step is to let $\alpha$ vary linearly with $T$. We leave that to future work.

\emph{(c) The distributed-order spectrum $w(\alpha)$.} Recovering $w$ from \eqref{eq:distorder} is a linear problem, but a severely ill-posed one: small errors in the data change the answer a great deal. We reconstruct $w$ with a non-negative, smooth fit. The weights are set through a softmax, so they stay positive and sum to one. For several conditions, the same fit is applied jointly across clocks, using the Caputo scaling law
\begin{equation}
\Dc^{\alpha}_t\big[v(kt)\big]=k^{\alpha}\,\big(\Dc^{\alpha}v\big)(kt),
\label{eq:scaling}
\end{equation}
Here $k=k(T)$ is the Arrhenius clock rate from ~\eqref{eq:Qg}. Under a change of temperature, each order $\alpha$ therefore leaves its own $k^{\alpha}$ signature. Even so, the fine shape of $w$ is not expected to be identifiable from realistic windows. The distinct power-law regimes only separate across time spans that a single aging record does not cover. Rescaling time across clocks does not add those spans (Section~\ref{sec:dynrange}).

\begin{figure}[t]
\begin{center}
\includegraphics[width=\linewidth]{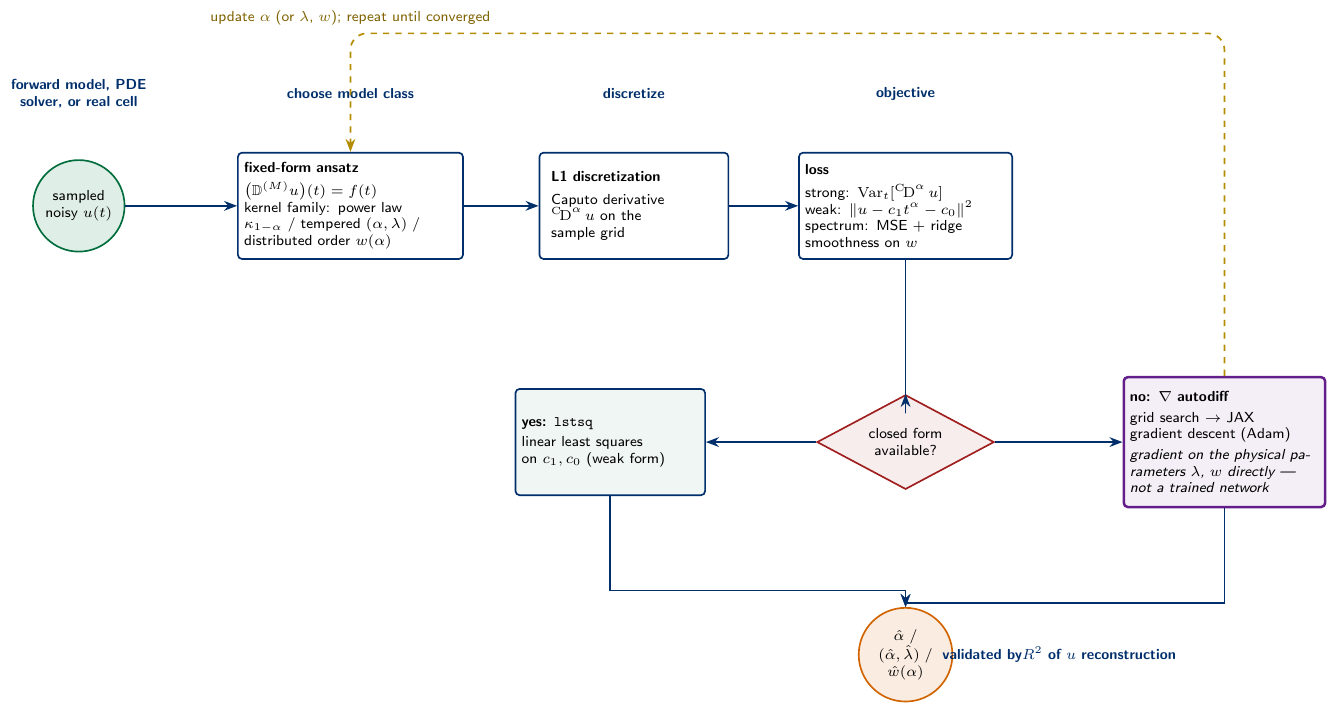}
\caption{Kernel-identification pipeline: ansatz, discretisation, loss, and the
closed-form/gradient-descent split, ending in the recovered order, tempered pair, or
spectrum, validated by the reconstruction $R^2$.}
\label{fig:optpipeline}
\end{center}
\end{figure}

Fig.~\ref{fig:optpipeline} shows the three estimators as one pipeline. A fixed-form model is discretised with the L1 scheme of ~\eqref{eq:L1}. The loss changes with the case: variance for the strong-form order, least squares for the weak form, and squared error plus a smoothness penalty for the spectrum. Only the tempered and spectrum cases need an iterative optimiser, because the weak-form order reduces to linear least squares.

\section{Results}\label{sec:3}
\setcounter{equation}{0}

\subsection{Non-power-law is detectable}

\begin{figure}[!ht]
\begin{center}
\includegraphics[width=\linewidth]{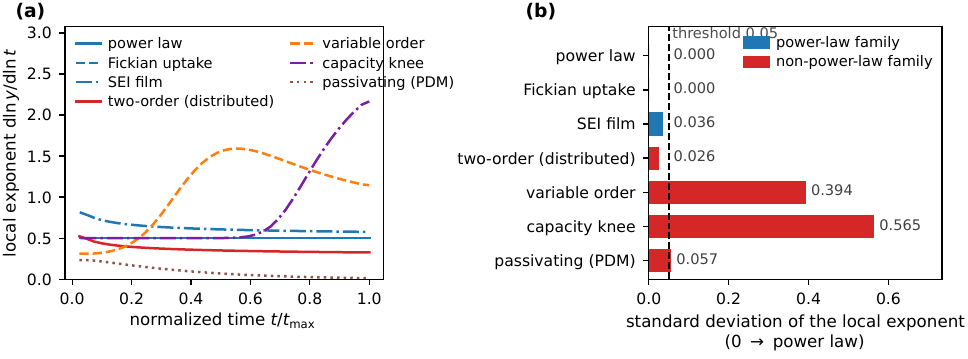}
\caption{Detecting non-power-law memory across the seven benchmark curves.
(a) Local log-log slope $\dd\ln y/\dd\ln t$: flat for the power-law family (blue, three
dash patterns) and varying for the non-power-law family (warm colours, one dash pattern
each). (b) Standard deviation of that local slope for each curve, against the $0.05$
threshold (dashed), with each value printed because two of the bars are shorter than a
line width; bar colour repeats the family coding of (a). The threshold separates the
strongly non-power-law curves but misclassifies the two-order (distributed) case at
$0.026$, which is caught instead by its power-law fit quality.}
\label{fig:diag}
\end{center}
\end{figure}

Slope variation is a useful but imperfect way to tell power-law curves from other curves. It measures how much the local log-log slope $\dd\ln u/\dd\ln t$ changes. The slope is constant for a power law and varies otherwise
(Fig.~\ref{fig:diag}). Power-law curves have slope variation below $0.05$, and a single order fits them with $R^2>0.999$. The strongly non-power-law curves reach $0.39$ and $0.57$. The boundary is not clean, though. The two-order (distributed) curve has slope variation $0.026$. That is below the $0.05$ threshold, and below the $0.036$ of a power-law member. The passivating curve sits just above the threshold, at $0.057$. So slope variation alone misclassifies the distributed case. We separate it by how well a power law fits, and we use both criteria together.

\subsection{The effective order can be recovered reliably}

\begin{figure}[!ht]
\begin{center}
\includegraphics[width=0.62\linewidth]{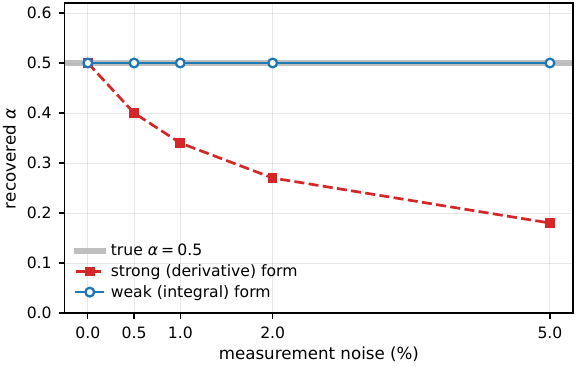}
\caption{Effective-order recovery versus noise on a single power-law trajectory. The
weak or integral form (open circles) returns $\alpha=0.500$ at every noise level tested
and therefore lies on the true-$\alpha$ reference (thick grey line); the strong
(derivative) form (filled squares) collapses. Markers sit at the five sampled noise
levels; the joining segments carry no information.}
\label{fig:order}
\end{center}
\end{figure}

For power-law data, the order is recovered almost exactly. With measurement noise, the two estimators behave very differently (Fig.~\ref{fig:order}). The weak or integral form keeps $\alpha=0.500$ up to $5\%$ noise. The derivative-matching strong form falls from $0.5$ to $0.18$. One lesson recurs throughout: avoid differentiating noisy data, which the strong form does here.

\subsection{Tempered parameters need multiple conditions}

\begin{figure}[!ht]
\begin{center}
\includegraphics[width=\linewidth]{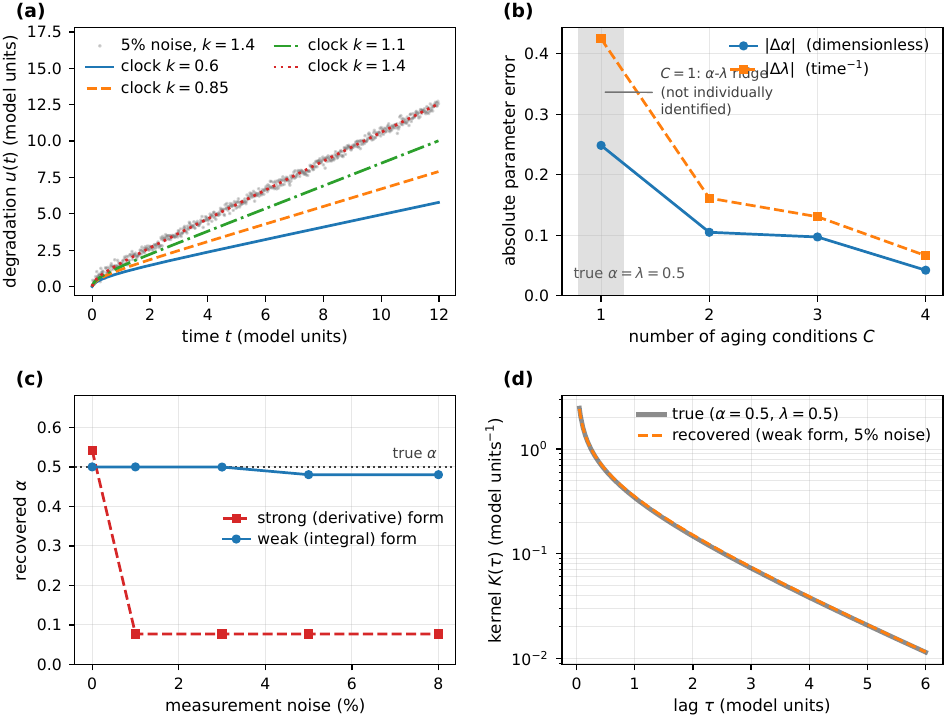}
\caption{Multi-condition identification of a tempered kernel; all quantities are in the
nondimensional units of the model problem. (a) One kernel seen through four Arrhenius
clocks (one dash pattern each), with the $5\%$-noise realisation of the fastest clock
$k=1.4$ shown as grey points behind the clean curves. (b) Absolute error in each
parameter as conditions are added, recovered by the strong form on clean data;
$|\Delta\alpha|$ is dimensionless and $|\Delta\lambda|$ has units of inverse time, and
they share an axis only because both true values are $0.5$. The shaded column at $N_{\mathrm{cond}}=1$
marks the regime discussed in the text, where the objective is a ridge and the plotted
point is the best-fitting combination rather than a localised estimate of either
parameter. (c) Recovery under noise for this four-condition fit: the weak form holds $\alpha$ at $0.500$ to $3\%$ noise and $0.480$ at $5$ and
$8\%$, while the strong form collapses after $1\%$. (d) The kernel recovered by the weak
form at $5\%$ noise against the true kernel.}
\label{fig:multi}
\end{center}
\end{figure}

For the tempered power-law kernel $M_{\alpha,\lambda}$ of ~\eqref{eq:tempered}, one curve does not determine $(\alpha,\lambda)$. The trouble is not just wide confidence intervals. A dense scan of the strong-form objective (step $0.005$ in both $\alpha$ and $\lambda$) shows that, at one condition, the minimum lies on a ridge. Along that ridge, $\alpha$ and $\lambda$ trade off against each other. Across nearly the whole search box, the objective stays within a factor of two of its best value. At $\lambda=0.005$, the best-fitting $\alpha$ is $0.78$, which is $1.01\times$ the minimum. At $\lambda=0.555$, it is $0.205$, which is $1.94\times$ the minimum. The true parameters $(0.5,0.5)$ score $18.4\times$ worse than the bottom of the ridge. The true parameters are not near the bottom of the objective, even on their own noiseless output. This is informative. The gap is not a flat direction around the truth. It may come from a bias in the strong-form objective. That objective weights the differentiated data, so it may penalise the true short-time behaviour more heavily than a shallower, earlier substitute. We have not tested this. It is another reason to prefer the weak form under noise. The multi-condition sweep that follows uses the same strong-form objective on clean data. So the comparison is fair: only the number of conditions changes, not the estimator. A single condition therefore constrains only a combination of $(\alpha,\lambda)$, not either parameter alone. Even the best point on the ridge, $(\hat\alpha,\hat\lambda)=(0.76,0.05)$, is far from the truth: $(|\Delta\alpha|,|\Delta\lambda|)=(0.25,0.43)$. Fitting one shared kernel across several Arrhenius-clock conditions removes the ridge and makes the parameters identifiable (Fig.~\ref{fig:multi}). The absolute error falls to $(0.10,0.16)$ at two conditions and to $(0.04,0.07)$ at four. The weak form keeps this accuracy under noise.

\subsection{Mechanistic and analytical cross-validation}
\label{sec:3.4}

\begin{figure}[!ht]
\begin{center}
\includegraphics[width=\linewidth]{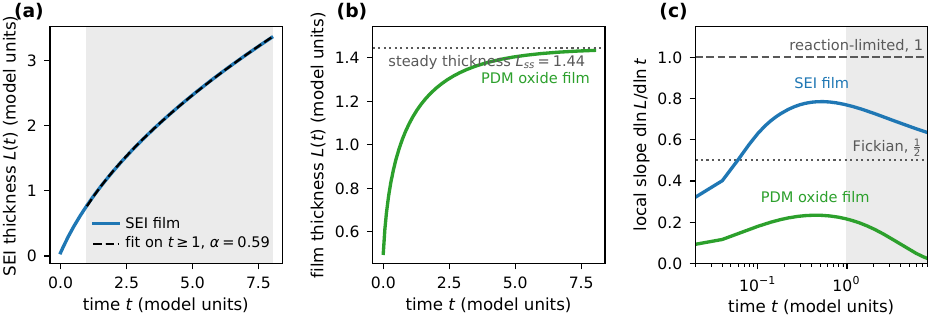}
\caption{Mechanistic data (Jacobian-free Newton--Krylov) identified by \texttt{gfckit};
all quantities are in the nondimensional units of the models. (a) SEI
reaction-diffusion thickness $L(t)$ with the weak-form power law of effective order
$\alpha=0.59$, drawn only over the window $t\geq1$ on which it was identified (shaded).
(b) Passivating corrosion film from the Point Defect Model, approaching its steady
thickness $L_{ss}=1.44$ (dotted). (c) The local growth-law slope $\dd\ln L/\dd\ln t$ of
both models, against the reaction-limited ($1$, dashed) and Fickian ($\tfrac12$, dotted)
references. Over the fit window the SEI slope falls monotonically from $0.77$ to $0.64$,
which is the trend toward $\tfrac12$ referred to in the text; the PDM slope falls toward
zero as the film saturates. The pointwise local slope and the offset-absorbing
weak-form order $\alpha=0.59$ are different summaries of the same window.}
\label{fig:phys}
\end{center}
\end{figure}

We feed the mechanistic curves from the Newton--Krylov solver into the identifier. This is a cross-model test. The generators contain no fractional operator, so nothing guarantees that the fit gives a sensible result (Fig.~\ref{fig:phys}). The SEI reaction-diffusion model gives an effective exponent $\alpha\approx0.59$. This is the weak-form estimate, and it drifts by about $\pm0.02$ across the sampled window and noise realisations. The value lies between the reaction-limited value ($1$) and the diffusion-limited value ($\tfrac12$), and it moves towards $\tfrac12$. Fickian diffusion gives $\alpha=0.500$ exactly. The PDM curve saturates.

Both mechanistic generators, the SEI film and the PDM oxide, are our own models. Agreement with them shows internal consistency, but it cannot rule out that the method has been checked only against itself. We therefore add a third test. Its answer is fixed in advance by classical electrochemistry.
The de Levie cylindrical pore~\cite{DeLevie1963} is an $RC$ transmission line. The resistance $R$ is per unit length along the pore, and the double-layer capacitance $C$ is per unit length across its wall. The model has a closed-form solution,
\begin{equation}
Z(\omega) \;=\; \sqrt{\frac{R}{j\omega C}}\;
              \coth\!\left(\ell\sqrt{j\omega R C}\right),
\label{eq:delevie}
\end{equation}
Here $\ell$ is the pore length. Above the crossover $\omega_c = 1/(RC\ell^{2})$, the hyperbolic cotangent is close to one, and $Z \to \sqrt{R/C}\,(j\omega)^{-1/2}$. This is a constant-phase element with exponent exactly $\tfrac12$, whatever the parameters. Below $\omega_c$ the pore behaves like a single capacitor, with exponent $1$. So the target value is known in advance. We read the order from the high-frequency branch with a log-log slope estimator. It is the frequency-domain version of the weak form in Section~\ref{sec:2.7}. Changing the domain is the only concession this test needs.
The method returns $\alpha = 0.500000$ (root-mean-square, RMS, fit error $6.6\times10^{-8}$; mean phase $-45.0000^{\circ}$). With $1\%$ relative noise added, it gives $\alpha = 0.4998$. The fitted amplitude matches $\sqrt{R/C}$ to six figures. This agreement is with a target set by an outside formula, not with another of our simulations.

\begin{figure}[!ht]
\begin{center}
\includegraphics[width=\linewidth]{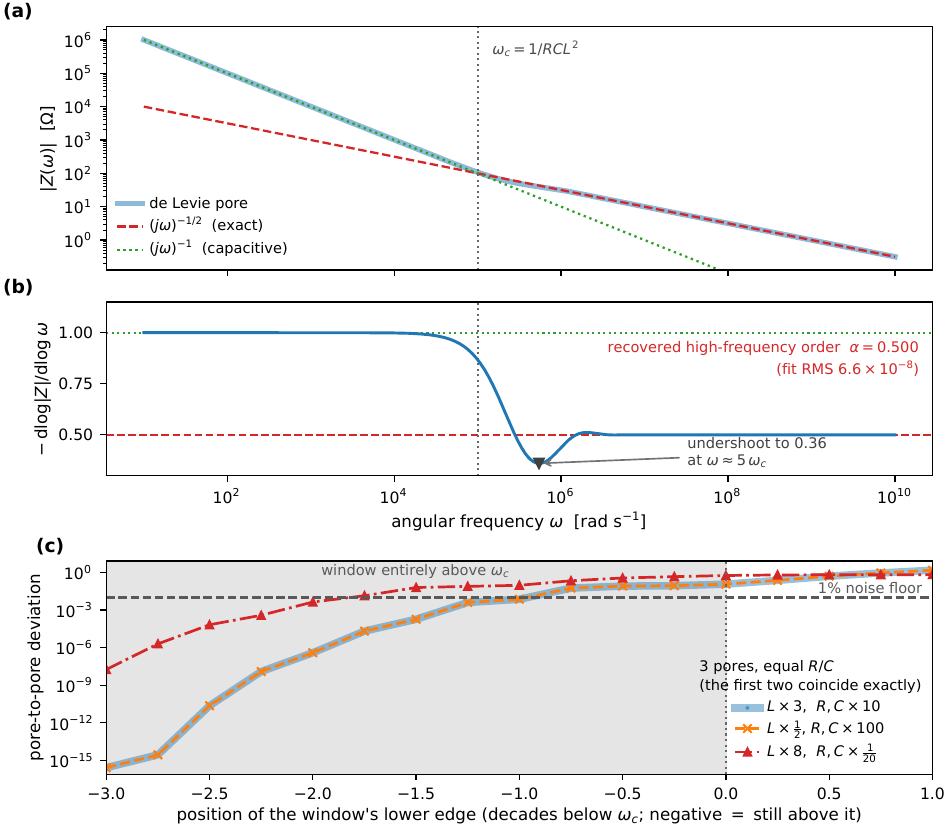}
\caption{External analytical validation on de Levie's porous electrode
\eqref{eq:delevie}. (a) The closed-form impedance with its two exact asymptotes,
$(j\omega)^{-1/2}$ above the crossover $\omega_c=1/(RC\ell^{2})$ and $(j\omega)^{-1}$ below
it. (b) The local exponent, recovered as $\alpha=0.500$ in the high-frequency branch;
note the undershoot to $0.360$ at $\omega\approx5\,\omega_c$. (c) Three pores of
different length and different $R$, $C$ but equal $R/C$: indistinguishable at machine
precision once the window lies about three decades above $\omega_c$, and separable
above a $1\%$ floor only once its lower edge comes within one to two decades of the
crossover. The abscissa of (c) counts decades \emph{below} $\omega_c$, so its negative
half is a window still entirely above the crossover (shaded). Two of the three curves
coincide exactly and are drawn as a wide pale line under a marked thin one; see
Section~\ref{sec:3.4}.}
\label{fig:delevie}
\end{center}
\end{figure}

The same formula also shows why the negative result of Section~\ref{sec:3.5} happens (Fig.~\ref{fig:delevie}). In the high-frequency branch, the response depends on $R$ and $C$ \emph{only} through $\sqrt{R/C}$. It does not depend on $\ell$ at all. Pore length is not weakly constrained there. It is absent from the formula. Take pores whose lengths range from one half to eight times the reference, and whose $R$ and $C$ range from one twentieth to a hundred times it. The length spread is $16$, and the $R$, $C$ spread is $2000$, between the extremes. These pores share $R/C$, so their responses agree to $2.6\times10^{-16}$ (machine precision) once the window lies about three decades above the reference pore's crossover $\omega_c$. They agree to better than $10^{-11}$ two and a half decades above it. They separate above a $1\%$ floor only when the window reaches to within one or two decades of the crossover. Only there does the geometry affect the response. So the placement of the window, not the flexibility of the model, decides what can be identified.

The same curve gives one practical warning. Between the two branches, the local exponent dips to $0.360$ near $\omega \approx 5\,\omega_c$ before it settles on $\tfrac12$. A narrow window placed at the crossover would report an effective order below both analytic limits, and it would still fit a power law well. So the position of the window is part of the measurement. An effective order given without its window is incomplete.

\subsection{The order spectrum is not identifiable, except for passivation}
\label{sec:3.5}

\begin{figure}[!ht]
\begin{center}
\includegraphics[width=\linewidth]{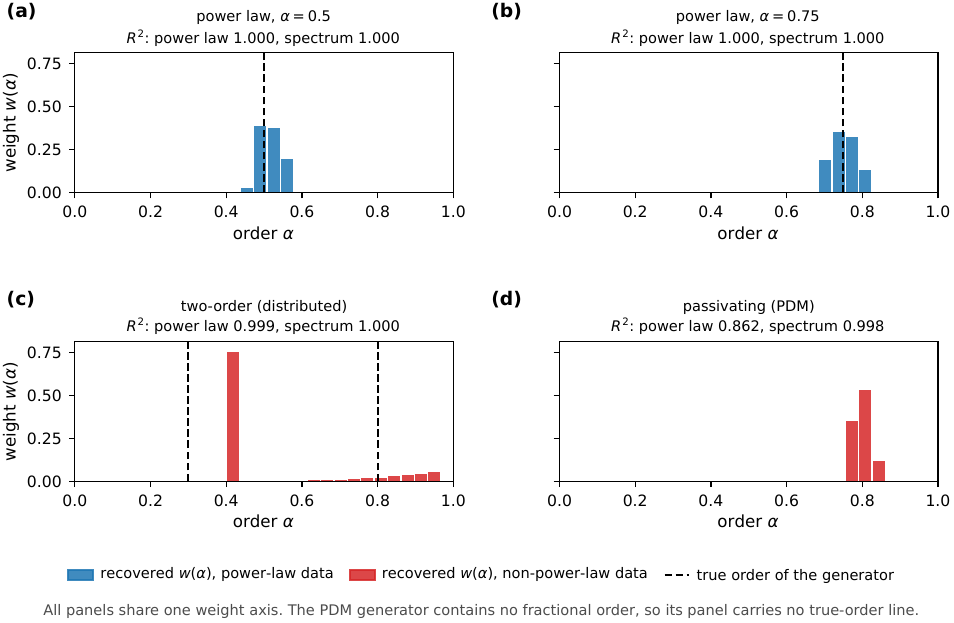}
\caption{Recovered order spectra $w(\alpha)$ for four benchmark curves, all drawn on one
weight axis so the panels are comparable. (a), (b) Power-law data give a single spike at
the true order (dashed). (c) The two-order (distributed) memory $\{0.3,0.8\}$ does not
give a bimodal spectrum: the mass collapses to a single intermediate order
$\alpha\approx0.42$, and the spectrum's own $R^2$ exceeds that of the single power law it
generalises, so the collapse is a property of the objective and not of the optimiser.
(d) Passivating PDM data, where the general kernel achieves the higher $R^2$ ($0.862$ to
$0.998$); the PDM generator contains no fractional order, so this panel has no true-order
line. The time-varying-order benchmark is fitted in the released code but is not shown
here: its response grows faster than any fixed mixture of orders in $[0.1,0.95]$ can
reproduce, so its low $R^2$ measures that model mismatch rather than identifiability, and
by the convergence diagnostic stated in the text no inference may be drawn from it.}
\label{fig:spec}
\end{center}
\end{figure}

Fig.~\ref{fig:spec} shows the central negative result. A distributed order is known to be unique under suitable conditions, even in an unknown medium~\cite{JinKian2023}. What follows is different. It asks how much data is needed to recover the order. Power-law data give a sharp single spike at the true order. A two-order memory $\{0.3,0.8\}$ with weights $\{0.6,0.4\}$ does not give two peaks. Instead, the recovered mass gathers near one intermediate order, $\alpha\approx0.42$, spread over a few neighbouring bins. A flexible single power law fits the same curve with $R^2=0.9989$.

This takes care. The fit converges slowly, and a fit that has not converged can look like the same result for the wrong reason. After $5{,}000$ optimiser steps the spectrum fit reaches only $R^2=0.84$. That is \emph{worse} than the single power law it contains as a special case. A richer model that fits worse than a simpler model it contains shows that the optimiser has failed. It does not show that the problem is ill-posed, so we draw no conclusion from it. We therefore run until the fit converges. At $5\times10^4$ steps it reaches $R^2=0.99997$, and it stays at $0.99998$ at $8\times10^4$. The smoothness penalty is not the limit. The results are the same for ridge weights of $10^{-5}$, $10^{-7}$ and $0$.

A second check is needed. If the method could never return two peaks, the collapse would look the same, for a reason unrelated to identifiability. So we give the estimator the easiest bimodal target we can: orders $\{0.2,0.9\}$ with equal weights, no noise, and $2\times10^5$ steps. It finds two peaks. Over $2.90$ decades, the recovery is $\{0.42,0.95\}$ with weights $\{0.63,0.31\}$. Over $3.20$ decades, it is $\{0.46,0.93\}$ with weights $\{0.75,0.18\}$. In both cases the upper order is within $0.05$ of the truth. So the estimator can return more than one peak, and the collapse of the $\{0.3,0.8\}$ benchmark is not caused by the parameterisation. The control also shows that finding two peaks is not the same as recovering the spectrum. The lower order lands at $0.42$ to $0.46$, while the true value is $0.2$. The $1$-Wasserstein gap, a measure of how far the mass has moved along the order axis, is $21\%$ and $27\%$ of that axis. This is close to the $27.9\%$ of the constructive bound in Section~\ref{sec:dynrange}. Even when conditions favour the method, and the orders are further apart than in the benchmark, the fit does not recover the generating spectrum at the window widths a single record provides. The control also shows the convergence trap again. At $5\times10^4$ steps, the $3.20$-decade case returns one mode with $R^2=0.9957$. That residual is more than two orders of magnitude worse than the converged fit. Running the optimiser longer reverses this verdict.

Once the fit has converged, the result is more than a failure to resolve. Take the forward model at the \emph{true} spectrum, on the same grid. Use the sink rate and source term $(\mu,S_0)$ of the relaxation ordinary differential equation (ODE), which differ from the kernel tempering rate $\lambda$ in ~\eqref{eq:tempered}. This gives $R^2=0.9872$. The recovered collapsed spectrum gives $R^2=0.99997$. So a single intermediate order reproduces the data better than the two-order memory that made them. The optimiser is not failing to find the truth. The objective prefers a simpler, wrong answer, and longer optimisation did not change that for the objectives and parameterisations we tried.

\begin{figure}[t]
\begin{center}
\includegraphics[width=\linewidth]{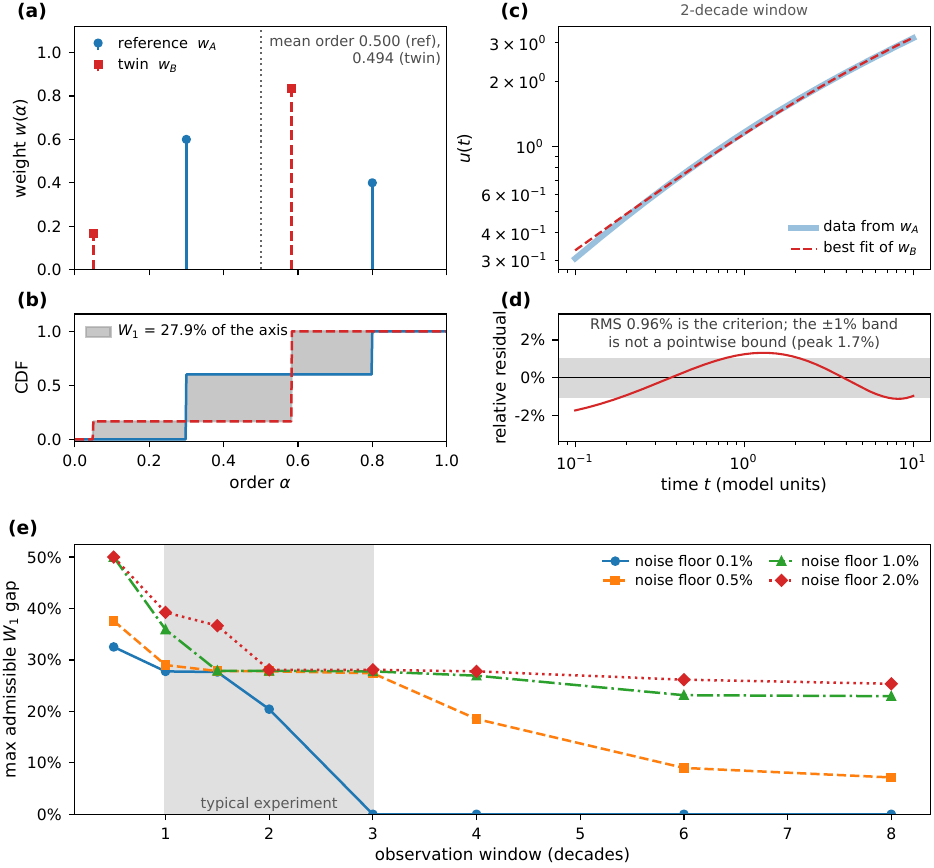}
\caption{Constructive non-identifiability. (a) The reference spectrum
$\{0.3,0.8\}$, weights $\{0.6,0.4\}$, and the farthest spectrum indistinguishable from it
over two decades at a $1\%$ noise floor; their mean orders are $0.500$ and $0.494$, the
one quantity the data constrain. (b) The $1$-Wasserstein gap drawn as the area between
the two cumulative distributions. (c) The two responses after the free amplitude and
offset any real fit absorbs. (d) The residual between them. The criterion is an RMS, not
a pointwise bound: the relative residual has RMS $0.96\%$, below the floor, while
leaving the $\pm1\%$ band (shaded) at both ends of the window, with a pointwise peak of
$1.7\%$. (e) The gap surviving as a function of window width and noise floor, with the
$1$ to $3$ decades shaded: the gap vanishes only when window and
precision are favourable together.}
\label{fig:keystone}
\end{center}
\end{figure}

One example shows only that recovery \emph{can} fail. The failure can also be built on purpose, which makes the point sharper. Keep the reference spectrum above. On a dense grid of two-order spectra over $\alpha\in[0.05,0.95]$, find the one farthest from it whose response the data cannot tell apart. Distance is the $1$-Wasserstein metric, scaled by the width of the axis. It is the fraction of the axis over which the mass has moved.
Two responses count as inseparable when their relative RMS difference is below a stated noise floor. We first remove the free amplitude and offset that any real fit absorbs. Only the shape of the curve, not its scale, then carries information. Over a two-decade window at a
$1\%$ noise floor the farthest such spectrum is $\{0.050,0.583\}$ with weights
$\{\tfrac16,\tfrac56\}$: it matches the reference data to $0.96\%$ RMS, yet it lies $27.9\%$ of the order axis away (Fig.~\ref{fig:keystone}). The two spectra share only one thing that the data constrain: their mean order, $0.500$ against $0.494$. They differ in everything else. A gap this large, below the noise floor, shows that no simple curve-fitting method can tell them apart at that precision and duration.

Varying the window width and the noise floor separates two requirements that are easy to confuse (Fig.~\ref{fig:keystone}e). A wider range alone does not make the spectrum identifiable:
at a $1\%$ noise floor a $23.0\%$ gap survives even at eight decades, so extending the
experiment without improving precision leaves the spectrum nearly as ambiguous as before.
Better precision alone does not help either: at a $0.1\%$ floor, the gap is still $20.4\%$ at two decades. Only both together close the gap. At a $0.1\%$ floor, the admissible gap reaches zero by three decades. The reference then becomes the only spectrum consistent with its own data on the searched grid. So identifiability needs progress on two fronts at once. Section~\ref{sec:dynrange} measures the first one, the range of time the two regimes need in order to separate.

The exception is passivation that saturates fully. A curve that slows down but keeps growing is not this case. The PDM film saturates. A single power law cannot represent that ($R^2=0.862$), but the general (distributed) kernel does, with $R^2=0.998$. The recovered spectrum does not have two peaks. It is one sharp intermediate order, $\alpha\approx0.81$, and $99.998\%$ of the spectral mass sits in three neighbouring bins. The general kernel can represent saturation because of the ODE form that goes with that order. A power law cannot saturate at all. So self-limiting degradation is the case where a general fractional operator is needed.

\subsection{Quantifying the boundary: dynamic range}\label{sec:dynrange}

\begin{figure}[!ht]
\begin{center}
\includegraphics[width=\linewidth]{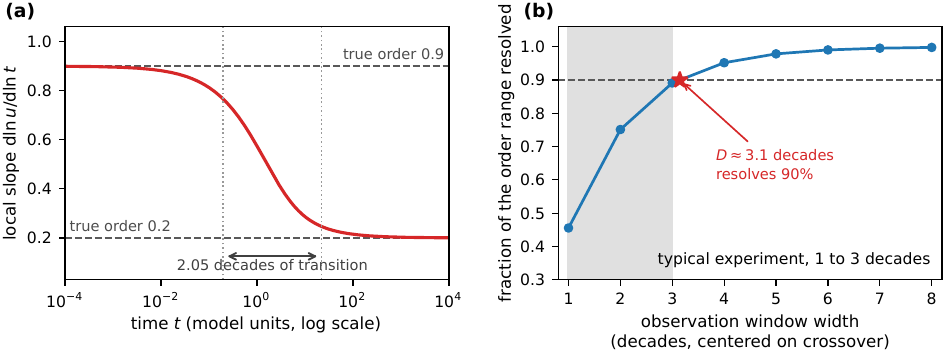}
\caption{Dynamic-range requirement for spectrum identifiability (fixed-Talbot
inversion). (a) The local slope reaches the true orders $0.2$ and $0.9$ (dashed) only far
from the crossover; the marked interval is the $2.05$ decades of transition between them.
(b) Fraction of the order range resolved versus observation-window width: $0.46$, $0.75$
and $0.89$ at one, two and three decades, with $D\approx3.1$ (star) needed for $90\%$.
The shaded band is the $1$ to $3$ decades shown, the same band as in
Fig.~\ref{fig:keystone}e.}
\label{fig:dr}
\end{center}
\end{figure}

Why can the spectrum not be resolved? A distributed-order memory has one power-law regime for each order. The regimes meet at a crossover time, and the local slope moves smoothly between the orders across that crossover. Using the fixed-Talbot inverse Laplace transform, we compute the slope over eight decades of time for a well-separated pair, $\{0.2,0.9\}$ (Fig.~\ref{fig:dr}a). The slope approaches each order very slowly. It is near $0.9$ only for $t\lesssim0.2$, and near $0.2$ only for $t\gtrsim22$. The two regimes are therefore separated by about two decades of transition ($\log_{10}(22.2/0.198)=2.05$). During that transition, the data look like one intermediate order.
To measure how much of the order range $[0.2,0.9]$ a window resolves, we use a window of width $D$ decades, centred on the crossover (Fig.~\ref{fig:dr}b). Resolving $90\%$ needs $D\approx3.1$ decades. A one-decade window resolves $0.46$ of the range, two decades resolve $0.75$, and three decades resolve $0.89$. This is the best case, with equal weights, widely separated orders, and clean, dense data. Closely spaced orders, noise and sparse sampling all raise the requirement a lot. Changing the clock across conditions rescales time but adds no decades, so it cannot help. This matches the argument above. A multi-condition experiment should also leave the recovered spectrum collapsed.

\subsection{The boundary measured on published cells}\label{sec:3.7}

The $3.1$-decade requirement describes experiments, not our generators. So it can be checked on real cells without knowing any cell's true values. We use twenty cells from two independent public datasets. The Oxford Battery Degradation Dataset~\cite{Birkl2017} has eight cells, checked with periodic pseudo open-circuit-voltage (OCV) tests. Zhang et~al.~\cite{Zhang2020} have twelve cells, whose capacity is logged every cycle at three temperatures. For each cell we measure two things, and neither needs a kernel inversion. One is the number of decades its usable observation window spans. The other is the effective order of its capacity-loss curve, from a single power-law fit. We count decades in cycle number. The count runs from the first check where the capacity loss is clearly above zero to the last recorded one. This starting edge makes the usable window shorter than the whole test.

\begin{figure}[!ht]
\begin{center}
\includegraphics[width=\linewidth]{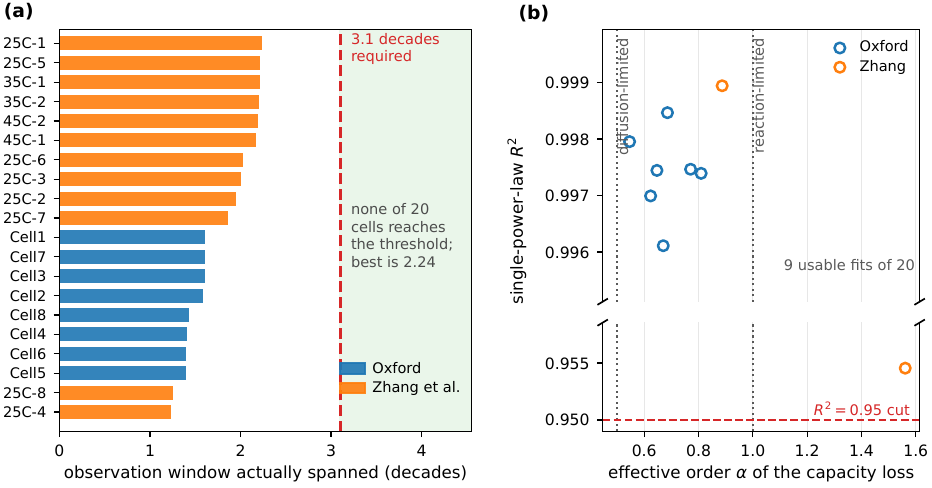}
\caption{The boundary of Section~\ref{sec:dynrange} measured on twenty published cells.
(a) The observation window each cell actually spans, against the $3.1$ decades required
for $90\%$ resolution: none reaches it and the best falls $0.86$ decades short.
Bars are coloured by dataset (blue, Oxford~\cite{Birkl2017}; orange, Zhang
et~al.~\cite{Zhang2020}), and the green region marks windows long enough to satisfy the
requirement, which none reaches.
(b) The effective order recovered from the same curves, which \emph{is} identifiable:
the Oxford cells lie between the diffusion- and reaction-limited values at median
$R^2=0.997$, and the two surviving Zhang et~al.\ fits (one within that
range, one above it) are also shown. The $R^2$ axis is broken so that the spread among
eight of the nine fits above $0.996$ is visible alongside the single fit at the $0.95$ cut. Fits
pinned to a search bound or with $R^2\le0.95$ are excluded and counted.}
\label{fig:realwin}
\end{center}
\end{figure}

Every cell shows the same result (Fig.~\ref{fig:realwin}a). Measured windows span $1.23$ to $2.24$
decades, with a median of $1.74$; \emph{none} of the twenty reaches $3.1$, and the
widest falls $0.86$ decades short. The shortfall is built into the test, not a chance result. A cell has to age a lot before any fade can be measured. That sets the start of the window. The end of the window is set by end of life. So a longer test adds far fewer decades than its length suggests. Section~\ref{sec:dynrange} shows that adding temperatures cannot make up the missing decades. The evidence shows that the boundary is not just a warning about ideal data. It lies beyond what a standard degradation test delivers.

The same cells do give an effective order (Fig.~\ref{fig:realwin}b). For seven of the eight Oxford cells, a single power law describes the loss curve well, with median $R^2=0.997$. The fitted $\alpha$ lies between $0.55$ and $0.81$. This is the same range as the SEI model in Section~\ref{sec:3}, which lies between the diffusion-limited value $\tfrac12$ and the reaction-limited value $1$. The two remaining Zhang fits give $0.89$ and $1.56$. The second lies outside both mechanistic limits, and its $R^2$ is at the cut-off. For the nine fits reported, the identifiability map also holds for real cells. What the map calls recoverable is recovered from real data. The window it says is needed for the spectrum is not available in either public dataset examined here.

This comparison has two limits. Eleven of the twenty order fits are discarded. Either the best fit lands on the edge of the search grid, or a single power law does not describe the curve ($R^2\le0.95$). Ten of these are Zhang cells, and one is an Oxford cell. The Zhang capacity is logged every cycle, so it is much noisier than periodic check-up data. Several of those cells fail at a knee rather than fading gradually. We also do not \emph{report} a recovered spectrum from these curves. On the Oxford cells, the spectrum fit converges to a worse result than its own nested single power-law case. By the test in Section~\ref{sec:3.5}, that points to an optimiser failure, not to ill-posedness. So no conclusion can be drawn from it either way. The dynamic-range argument needs no such fit, and that is why the comparison rests on it.

\section{Discussion: the identifiability map}\label{sec:4}
\setcounter{equation}{0}

Table~\ref{tab:map} summarises what can and cannot be recovered. Three practical points follow. First, reports that fit distributed-order or free-kernel models to single aging curves should be read with care. The extra freedom in the kernel is largely unconstrained by the data. A single effective order, plus a tempering scale where needed, is what the data support. Second, experimental design matters more than model flexibility. Identifiability comes from time range, measurement precision and multiple conditions, not from richer kernels. These are separate requirements, and neither can replace the other. At a $1\%$ noise floor, a $23\%$ spectral gap survives eight decades. At a $0.1\%$ floor, three decades close it completely. A longer experiment without better precision therefore buys much less than its length suggests. The window must also cover the decade where the quantity of interest is encoded. For the porous electrode, pore geometry appears in the response only near $\omega_c$ and is absent above it. A sweep of any width in the high-frequency range therefore says nothing about the geometry. Third, the boundary is operative, not asymptotic. Across twenty cells from two public datasets (Section~\ref{sec:3.7}), no observation window exceeds $2.24$ decades. The regime where a spectrum could be resolved is therefore outside what these standard protocols reach. A study that reports a recovered spectrum from such data is reporting a regularisation choice, not a measurement.

The $3.1$-decade figure is best read as a measure of how ill-posed this inverse problem is. It is stated in a quantity the experimenter controls, not as a convergence rate. The analytic literature shows that the order, the multi-term orders and the distributed order are in principle determined by the data, sometimes even when the medium is unknown~\cite{Cheng2009,LiuYamamoto2023,JinKian2023}. It also bounds stability in terms of the forward map~\cite{ZhangZhou2023,CenJinLiuZhou2023}. Neither result tells a practitioner whether a two-decade protocol is enough. The criterion here fills that gap for one class of kernels, and it can be checked against a sampling record before any inversion. Its limitation is that it rests on a benchmark rather than a proof. It is therefore only as general as the kernel families and observation model of Section~\ref{sec:2}. An analytic counterpart is the natural next step.

\begin{table}[!ht]
\begin{center}
\caption{Identifiability map: what data-driven general-fractional recovery does and does
not support.}
\label{tab:map}
\begin{tabular}{@{}p{0.475\linewidth}p{0.475\linewidth}@{}}
\toprule
\textbf{Recoverable} & \textbf{Not recoverable} \\
\midrule
Non-power-law detection (local-slope variation) & Fine-grained order spectrum
$w(\alpha)$ from a single curve \\
Effective order $\alpha$ (weak form; robust to $5\%$ noise) & The same spectrum from
multiple Arrhenius conditions, which rescale time without adding decades \\
Tempered $(\alpha,\lambda)$ via multi-condition fitting & Pore geometry from the
high-frequency branch of a porous electrode: $L$ is absent from the response and $R$,
$C$ enter only as $\sqrt{R/C}$ \\
Passivation signature (general kernel, $R^2$ from $0.862$ to $0.998$) & The decomposition
behind that fit: the recovered spectrum is one sharply resolved intermediate order, not a
separation into fast and slow modes \\
Exponents fixed by an external closed form (de Levie, $\alpha=0.500$ against an exact
$\tfrac12$) & The spectrum from published cells~\cite{Birkl2017,Zhang2020}: no cell spans more than $2.24$ decades, against the $3.1$ needed \\
\bottomrule
\end{tabular}
\end{center}
\end{table}

\section{Conclusion}\label{sec:5}
\setcounter{equation}{0}

The main practical result is a number. Resolving a distributed order to $90\%$ needs $3.1$ decades of observation. An experiment that falls short should report an effective order, not a spectrum. General fractional calculus suits memory-driven degradation, and recovering its kernel from data has limits that can be checked in advance. From realistic aging data one can recover an effective anomalous order. With several conditions, one can also recover tempered parameters and a clear sign of passivation. A fine-grained fractional-order spectrum cannot be recovered. These limits were mapped with synthetic generators, two independent mechanistic models, and twenty cells from two public datasets. Eleven order fits were discarded, because they reached a search bound or a single power law described them poorly. Broader tests across chemistries and degradation modes remain open. The generators, identification methods, labelled benchmark and all figures are openly available at \texttt{github.com/Feugmo-Group/gfckit}, so that anyone can test and extend these limits.

\section*{Acknowledgments}
C.G.T.F.\ acknowledges the support of the Natural Sciences and Engineering
Research Council of Canada (NSERC) [RGPIN-2024-03989]. This research was
enabled in part by support provided by SHARCNET and the Digital Research
Alliance of Canada.

\section*{Data availability statement}
The generators, identification methods, labeled benchmark, and figure scripts
comprising the \texttt{gfckit} package are openly available at \texttt{github.com/Feugmo-Group/gfckit}. The two measured datasets used in Section~\ref{sec:3.7} are
public and available from their original sources: the Oxford Battery
Degradation Dataset~1 ~\cite{Birkl2017} and the data of Zhang
et~al.~\cite{Zhang2020} (Zenodo 3633835).

\bibliographystyle{plainnat}
\bibliography{refs}

\end{document}